\documentclass[aps,
twocolumn,
showpacs,preprintnumbers,amsmath,amssymb,prx,floatfix,10pt]{revtex4-2}
\usepackage{graphicx}
\usepackage{mathtools}
\usepackage{bbm}
\usepackage[normalem]{ulem}
\usepackage{soul}
\usepackage[usenames,dvipsnames]{xcolor}
\usepackage{comment}
\usepackage{amsthm} 
\usepackage{amssymb}    
\usepackage[colorlinks,breaklinks=true,linkcolor=blue, citecolor=blue]{hyperref}

\begin{document}
\title{Apparent inconsistency between Streda formula and Hall conductivity in reentrant integer quantum anomalous Hall effect in  twisted MoTe$_2$}

\author{Yi Huang}
\thanks{These authors contributed equally to this work}
\author{Seth Musser}
\thanks{These authors contributed equally to this work}
\author{Jihang Zhu}
\author{Yang-Zhi Chou}
\author{Sankar Das Sarma}
\affiliation{Condensed Matter Theory Center and Joint Quantum Institute, Department of Physics, University of Maryland, College Park, Maryland 20742, USA}
\begin{abstract}
Recent experiments in twisted bilayer MoTe$_2$ (tMoTe$_2$) have uncovered a rich landscape of correlated phases. In this work, we investigate the reentrant integer quantum anomalous Hall (RIQAH) states reported by F. Xu \textit{et al.}, \href{https://doi.org/10.48550/arXiv.2504.06972}{arXiv:2504.06972 [cond-mat]} which display a notable mismatch between the Hall conductivity measured via transport and that inferred from the Streda formula. We argue that this discrepancy can be explained if the RIQAH state is a quantum Hall bubble or Wigner crystal phase, analogous to similar well-established phenomena in two-dimensional (2D) GaAs quantum wells. While this explains the RIQAH state at filling $\nu = -0.63$, F. Xu \textit{et al.} report that the other RIQAH state at $\nu = -0.7$ has a smaller slope, necessitating a different interpretation. We propose and substantiate with analysis of the experimental data that this discrepancy arises due to a nearby resistive peak masking the true slope. Furthermore, we identify this resistive peak as a signature of a phase transition near $\nu = -0.75$, possibly driven by a van Hove singularity. The anomalous Hall response and Landau fan evolution across this transition suggest a change in Fermi-surface topology and a metallic phase with a non quantized Hall response. These observations offer new insights into the nature of the RIQAH states and raise the possibility that the nearby superconducting phase may have a valley-imbalanced metal parent state. 
\end{abstract}

\maketitle

\section{Introduction}

In recent years van der Waals materials have emerged as a powerful playground for studying a variety of strongly correlated phases. An exciting recent development is the observation of the fractional quantum Hall effect (FQHE) at zero field \cite{cai_signatures_2023, park_observation_2023, Xu:2023, zeng_thermodynamic_2023, lu_fractional_2024, Park:2025, xu_signatures_2025}, dubbed the fractional quantum anomalous Hall (FQAH) effect~\cite{Sun:2011,Sheng:2011,Tang:2011,Neupert:2011,Regnault:2011}. Additionally, the FQHE has also been observed in van der Waals materials under the application of a small but nonzero magnetic field \cite{spanton_observation_2018, xie_fractional_2021, aronson_displacement_2024}.

The first observation of FQAH physics occurred in twisted bilayer MoTe$_2$ (tMoTe$_2$) \cite{cai_signatures_2023, park_observation_2023, Xu:2023, zeng_thermodynamic_2023}. These experiments found evidence for the $2/3$ and $3/5$ FQAH states. As samples have gotten cleaner correlated states beyond these FQAH states have emerged \cite{Park:2025, xu_signatures_2025}. For example, a very recent work studying a sample of flux-grown \cite{disorder} tMoTe$_2$ at a twist angle of $3.8^{\circ}$ found additional FQAH plateaus. The same sample also displayed a putative superconducting (SC) state next to the $2/3$ FQAH state \cite{xu_signatures_2025}. There have been very recent suggestions that this SC is an example of an anyon SC \cite{shi_doping_2024, Shi:2025, Nosov:2025, pichler_microscopic_2025}, or originates either from a charge density wave (CDW) metal \cite{zhang_holon_2025} or from spin-valley polarized intravalley pairing \cite{xu_chiral_2025}.

\begin{figure*}[t]
    \centering
    \includegraphics[width=\textwidth]{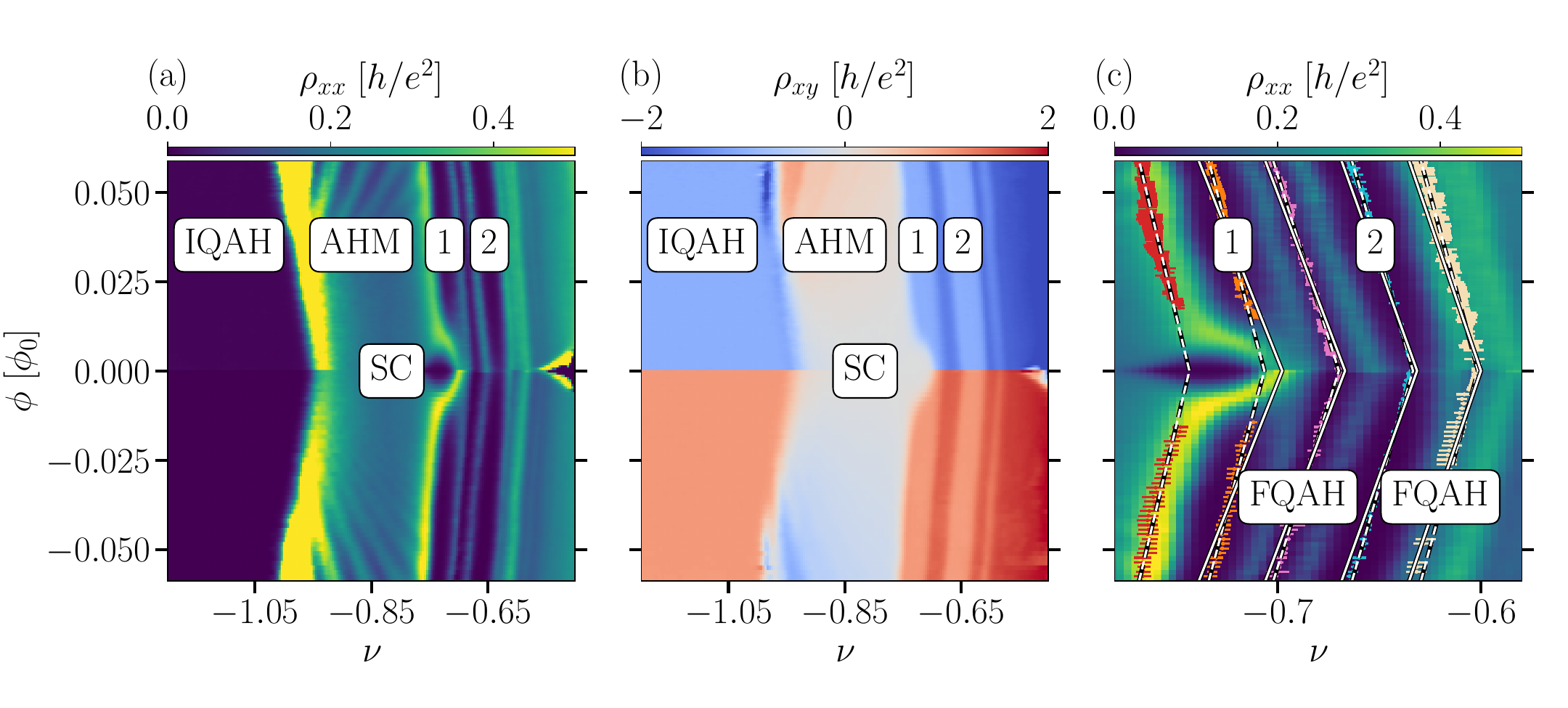}
    \caption{The experimental longitudinal and Hall resistivity, $\rho_{xx}$ and $\rho_{xy}$, respectively, of the sample from Ref.~\cite{xu_signatures_2025} are plotted in panels (a) and (b). Both are plotted as a function of magnetic flux per unit cell, $\phi$, and filling per unit cell, $\nu$. The vertical scale $\phi$ of all plots is identical, with tick marks provided for visual aid. Data are taken at a temperature of $15$mK and zero displacement field. We have added labels to a few features. The IQAH state has vanishing $\rho_{xx} \ll h/e^2$ and quantized $\rho_{xy} \approx h/e^2$. The anomalous Hall metal (AHM) appears across a resistive $\rho_{xx}$ peak. At small magnetic field it has larger $\rho_{xx}$ than the IQAH state and nearly zero $\rho_{xy} \ll \rho_{xx} \ll h/e^2$. At larger magnetic fields a Landau fan emanating from $\nu = -1$ becomes apparent. Feature (1) corresponds to the first reentrant integer quantum anomalous Hall (RIQAH1) state, characterized by low $\rho_{xx}$ and quantized $\rho_{xy} \approx h/e^2$. 
    This RIQAH1 state terminates before reaching $B = 0$ due to the emergence of SC, a putative superconducting phase. There is a second RIQAH state at feature (2), RIQAH2. In panel (c) the longitudinal resistivity is again plotted, but now with selected resistivity minima and maxima extracted. Appendix~\ref{app:line_fitting} details the process of extracting these minima and maxima and fitting to the white lines. Here solid white lines are fits where the slope is equal to the intercept, and dotted white lines are fits where they are not. We show RIQAH1 and RIQAH2, highlighted by the markers (1) and (2) and orange and cyan points, respectively. For reference we also show the FQAH states at $-2/3$ and $-3/5$, highlighted by the FQAH markers and pink and tan points, respectively. We find that both FQAH state and the RIQAH2 state are well fit by lines with intercepts equal to their slopes. This can be seen in the small difference between the dotted and solid fits. However, the RIQAH1 state is not, as shown by the large difference between the dotted and solid fit. We expect this is because the slope of the nearby resistivity peak, highlighted by red points, corresponds to $C=-0.49\pm 0.08$, which may obscure the slope of RIQAH1. For more details on the slopes of the fits see Table~\ref{tab:streda_and_transport}.}
    \label{fig:streda_one}
\end{figure*}

In this work we focus on a different puzzling feature of the data in Ref.~\cite{xu_signatures_2025} which is interesting in its own right, as well as potentially useful for shedding light on the nature of the putative SC. This is the appearance of what the authors dub the ``reentrant integer quantum anomalous Hall'' (RIQAH) states, by analogy with the well-known reentrant integer quantum Hall (RIQH) states first found in GaAs quantum wells~\cite{du_strongly_1999, lilly_evidence_1999,Cooper:1999, eisenstein_insulating_2002, li_observation_2010,Fu:2019}. The RIQAH states display nearly quantized Hall conductivity, $\sigma_{xy}\approx e^2/h$. Measuring the slope of the minima in resistivity in a plot of density, $n$, and magnetic field, $B$, gives another way to measure $\sigma_{xy}$ via the Streda formula \cite{Streda_1982}. We often refer to such a Streda slope as $C$ since it corresponds to the (quantized) Chern number in an incompressible IQAH state, although the slope may not be quantized in general. When $C$ is measured in this way for the RIQAH state, it displays a smaller value than $\sigma_{xy}$ as measured via transport. In a single gapped incompressible Hall state general arguments suggest that this should not happen and $\sigma_{xy}$ found via transport and the Streda formula should always agree \cite{hallbootstrap}. However, if the RIQAH phase shares the same origin as conventional RIQH phases in Landau level (LL) physics --- where localized compressible electron solids, such as bubble or Wigner crystal (WC) phases, form relative to a specific magnetic filling fraction (rather than the moir\'e filling per unit cell) --- then this may be allowed \cite{Fogler:1996,Koulakov:1996,du_strongly_1999,lilly_evidence_1999,Cooper:1999, eisenstein_insulating_2002,li_observation_2010,Fu:2019,Chen:2019,Csathy:2024}.
We find that this possibility can explain the RIQAH2 state labeled in Fig.~\ref{fig:streda_one}. However, due to the small slope of the RIQAH1 state, reported by Ref.~\cite{xu_signatures_2025}, another explanation is likely needed for this state. We give more details on these observations in Sec.~\ref{sec:mismatch}.

\begin{table}[h]
    \centering
    \begin{tabular}{|c||c|c|}
    \hline
         & Streda slope, $C$ & Transport, $\sigma_{xy}h/e^2$\\
    \hline \hline
    RIQAH1& $-0.70 \pm 0.06$ & $-1 \pm 0.01$\\
    FQAH, $\nu=-2/3$& $-0.67\pm 0.02$ & $-2/3 \pm 0.01$\\
    RIQAH2& $-0.63 \pm 0.02$ & $-1 \pm 0.01$\\
    FQAH, $\nu=-3/5$ & $-0.60 \pm 0.05$ & $-3/5 \pm 0.05$\\
    \hline
    \end{tabular}
    \caption{A table showing the mismatch between the Streda slopes, $C$, and the transport measurements, $\sigma_{xy}$, for the RIQAH states. The RIQAH1, FQAH at $-2/3$, RIQAH2, and FQAH at $-3/5$ phases are marked in orange, pink, cyan, and tan, respectively, in Fig.~\ref{fig:streda_one}(c). The value of the slope comes from the slope of fits displayed as solid lines passing through the corresponding points. The error for this slope is worse for RIQAH1 since it is a worse fit to this data, as discussed in the main text. For more information on these fits, see App.~\ref{app:line_fitting}. The transport values are taken from Ref.~\cite{xu_signatures_2025} with error estimates from the text.}
    \label{tab:streda_and_transport}
\end{table}

As Fig.~\ref{fig:streda_one} reveals, the RIQAH1 state extrapolating to $\nu = -0.7$ at zero field appears near the putative SC and thus is important to understand. To this end we note that it also always appears next to a resistive $\rho_{xx}$ peak, as can be seen in Fig.~\ref{fig:streda_one}. Since the RIQAH states are identified via minima in $\rho_{xx}$ \cite{Tingxin} such a proximate peak in resistivity could cause misidentification of the slope. 

We therefore propose that the RIQAH1 state can be understood just as the RIQAH2 state with its larger slope being masked by the $\rho_{xx}$ peak. We substantiate this by performing our own data analysis on the data from Ref.~\cite{xu_signatures_2025}. We find that the resistivity minima associated with the RIQAH1 state fit best to a line with intercept not equal to its slope, as seen in Fig.~\ref{fig:streda_one}(c). This line has a slope corresponding to $C=-0.46\pm 0.04$ and an intercept of around $\nu = -0.70$. We also find that the nearby resistivity maxima has a slope $C = -0.49\pm 0.08$, close to the slope for RIQAH1. This provides support for our claim that the nearby resistivity peak may be masking the true slope of the RIQAH1 state (e.g., via a sharp phase transition)~\cite{mask}. For a visual indication of what the true slope of the RIQAH1 state may look like, see the solid line through the orange points in Fig.~\ref{fig:streda_one}(c), further discussed in Table~\ref{tab:streda_and_transport}.

This, however, begs the question of what causes the resistive peak. We propose that it is a signature of a phase transition occurring around $\nu = -0.75$ that is driven by a van Hove singularity (VHS). This would explain a number of features in the data. First, we note that for fillings $\nu \in (-1, -0.75)$ and low magnetic fields the Hall resistivity is nearly zero ($\rho_{xy} \ll \rho_{xx} \ll h/e^2$), referred to as the anomalous Hall metal (AHM) in Ref.~\cite{xu_signatures_2025}, while on either side $\rho_{xy}$ is nearly $h/e^2$. This behavior is robust. A similar small $\rho_{xy}\ll h/e^2$, including a small $\rho_{xx}$ peak near $\nu = -0.75$, was previously observed in samples of tMoTe$_2$ with twist angles of $3.7^{\circ}$ and $3.9^{\circ}$ \cite{park_observation_2023,Park:2025}. One possible explanation is that the anomalous Hall metal has partial valley polarization, separating it from the fully valley-polarized phases at $\nu\lesssim-1$ and $\nu \gtrsim -0.75$ by a phase transition. Moreover, at large magnetic fields the anomalous Hall metal has the opposite sign of $\rho_{xy}$ and the opposite slope of its nondegenerate Landau fan emanating from $\nu = -1$. The transition near $\nu = -0.75$ out of the anomalous Hall metal should therefore be expected to accompany a VHS where a Fermi surface of opposite charge carriers emerges. Another hint that there may be a VHS near $\nu = -0.75$ comes from a small peak in the reflective magnetic circular dichroism data taken in a tMoTe$_2$ device near this twist angle \cite{cai_signatures_2023}. That data displayed a small peak in coercive field near $\nu = -0.75$, possibly arising because the VHS may enhance the interaction-induced spin gap. We give more details in Sec.~\ref{sec:VHS}, along with suggestions on how to test our predictions.

Finally, we note that if there is in fact a VHS driven transition at $\nu = -0.75$ this may offer insights into the nature of the pairing mechanism for the SC found near this filling. If this is the case, the normal state for SC is likely a valley imbalanced metallic state with nearly vanishing $\rho_{xy}$, or an intervalley coherent (IVC) state that breaks time-reversal symmetry \cite{fischer_theory_2024, guerci_topological_2024, Bi:2021, Xie:2024}. These normal states can become superconducting through BCS-like pairings \cite{chou_acoustic-phonon-mediated_2021,zhu_superconductivity_2025}, Kohn-Luttinger \cite{ghazaryan_unconventional_2021, jimeno-pozo_superconductivity_2023, guerci_topological_2024, qin_kohn-luttinger_2024}, antiferromagnetic spin fluctuations \cite{fischer_theory_2024}, etc. This scenario complements the proposed anyon SC \cite{shi_doping_2024, Shi:2025, Nosov:2025, pichler_microscopic_2025} and SC arising from spin-valley-polarized intravalley pairing \cite{geier_chiral_2024, chou_intravalley_2025, wang_chiral_2024, yang_topological_2024, parra-martinez_band_2025, christos_finite-momentum_2025, xu_chiral_2025,Dong:2025,Gil:2025,Jahin:2025,Qin:2024,Yoon:2025}.
We discuss this possibility, along with the outlook for future experiments in Sec.~\ref{sec:disc}.

\section{Mismatch between Streda and transport}
\label{sec:mismatch}

We begin by reviewing the derivation of the Streda formula for a single gapped incompressible Hall state. If a system has zero longitudinal conductivity $\sigma_{xx}=0$ and has Hall conductivity given by $\sigma_{xy}$, then its transport response to an electric field will be
\begin{equation}\label{eqn:transport}
J_i = \sigma_{xy}\varepsilon_{ij}E_j,
\end{equation}
where $\varepsilon_{ij}$ is the Levi-Civita symbol. For example, in an IQAH state with Chern number $C$, we have $\sigma_{xy}= e^2C/h$. Taking $\partial_i$ of this expression and using Maxwell's equations will reveal that the time derivative of density is proportional to the time derivative of the magnetic field \cite{hallbootstrap}. If this equation is integrated over time and divided by the area of the unit cell, then we arrive at the Streda formula per unit cell \cite{Streda_1982,lu2020filling}
\begin{equation}\label{eqn:Streda}
\nu(\phi) = \frac{h\sigma_{xy}}{e^2} \frac{\phi}{\phi_0} + \nu(\phi=0),
\end{equation}
where $\nu$ is the filling fraction per unit cell, $\phi$ is the magnetic flux through a unit cell, and $\phi_0=h/e$ is the magnetic flux quantum. This equation tells us that states with nonzero $\sigma_{xy}$ will require a change in density to maintain their gap when flux is threaded. Moreover, the slope of this density change is given by the Hall conductivity.

We now offer a comment on how the Streda formula is used in experiment. 
In transport measurements the minima of $\rho_{xx}$, where localization is the strongest, is used to indicate the presence of a gapped state. These minima are tracked as a function of carrier density, $n$, and magnetic field, $B$. The slope of the resulting line can then be extracted, giving the Hall conductivity. It is typical to see a peak in resistivity in between neighboring $\rho_{xx}$ minima which often (but not always) signals a plateau transition.

The derivation of Eq.~\eqref{eqn:Streda} from Eq.~\eqref{eqn:transport} seems to indicate that it is not possible for a quantum Hall state to display different values of $\sigma_{xy}$ measured through transport and Streda formula, which is the conventional wisdom. However, this expectation is violated in the RIQAH state reported in Ref.~\cite{xu_signatures_2025}, where the two methods yield different results.
A few explanations for this discrepancy have already been proposed. Ref.~\cite{Shi:2025} suggests that incomplete edge equilibration may be a contributing factor. Indeed, nonideal contacts may probe only some edge channels due to the lack of equilibration~\cite{Chklovskii:1992,Chklovskii:1993,Alphenaar:1990,Kouwenhoven:1990}, especially when the edge-channel structure is complicated, i.e., in $2/3$ FQH state~\cite{Kane:1994,Kane:1995a,Kane:1995b,Fradkin:1997}. Edge equilibration issues should vanish in sufficiently large samples and can be probed experimentally by varying contact configurations or by engineering edge disorder and geometry. 
Additionally, Ref.~\cite{Nosov:2025} mentions that the RIQAH states may be compressible so that the Streda slope is nonuniversal and depends on disorder strength. 
This possibility can be tested via measuring the Streda slope for a different sample of different disorder strength using transport or other thermodynamic compressibility measurements~\cite{Young:2023,zeng_thermodynamic_2023}.

\begin{figure}[t]
    \centering
    \includegraphics[width=0.75\linewidth]{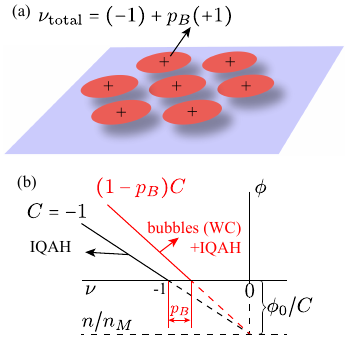}
    \caption{Schematics of the bubble or Wigner crystal phase + IQAH background. (a) The bubble or Wigner crystal phase involves electrons of a fixed fraction $p_B$ of the IQAH background that forms a lattice pinned by disorder. (b) Schematic of flux $\phi$ versus filling $\nu$. The lines indicate the center of $\rho_{xx}$ minima which shift as a function of $\phi$ and $\nu$ giving rise to the Streda slope. For a lattice system with an integer Chern band, the Berry curvature at $B = 0$ mimics a background magnetic field, which intercepts the conventional Landau fan at $1/C$ flux quantum, illustrated by the dashed lines.}
    \label{fig:CDW}
\end{figure}

We offer a different perspective by comparing the phenomenology of the RIQAH states to the conventional LL RIQH that appears at noninteger LL fillings $\nu_{\mathrm{mag}}$~\cite{Fogler:1996,Koulakov:1996,du_strongly_1999,lilly_evidence_1999,Cooper:1999, eisenstein_insulating_2002,li_observation_2010,Fu:2019,Chen:2019,Csathy:2024}. In that setting $\nu_{\mathrm{mag}}$ refers to the magnetic filling factor, rather than the density per moir\'e unit cell. It is well-known that the RIQH in both GaAs~\cite{du_strongly_1999,lilly_evidence_1999,Cooper:1999, li_observation_2010,eisenstein_insulating_2002,Fu:2019,Csathy:2024} and graphene~\cite{Chen:2019,Young:2023} appears at noninteger magnetic filling fractions which do not agree with the integer $\sigma_{xy}$ as measured by transport. Additionally, the RIQH states are observed at the same magnetic filling fraction even if the density, $n$, and magnetic field, $B$ are changed. Thus we expect the RIQH states should display exactly the same mismatch between Streda formula and transport Hall conductivity that the RIQAH states do.

As pointed out in Ref.~\cite{xu_signatures_2025}, the transport data in high-quality tMoTe$_2$ also bears a remarkable resemblance to the lowest LL (LLL) transport data in GaAs intentionally doped with Al reported in Ref.~\cite{li_observation_2010}. In that sample a RIQH state is observed between two FQH plateaus $\nu_{\mathrm{mag}}=2/3$ and $\nu_{\mathrm{mag}}=3/5$. This is very similar to the RIQAH2 state that intervenes between the FQAH states at $\nu=-2/3$ and $-3/5$.

The experiments \cite{li_observation_2010, xu_signatures_2025} are similar in another way; short-range disorder dominates both.
For the Al doped GaAs \cite{li_observation_2010} much weaker LLL reentrant behavior is observed for a sample with reduced short-range Al alloy disorder, and no LLL reentrant behavior is seen in clean samples without Al.
For transition metal dichalcogenide (TMD) materials, while the flux-grown synthesis method has significantly improved the crystal quality by reducing atomic point defects \cite{disorder}, transport measurements in these high-quality samples still show mobilities, $\mu$, that are proportional to $n^{-1/2}$ for $n \gtrsim 10^{12}$ cm$^{-2}$, confirming that short-range defects remain the dominant scattering source~\cite{CDean_WSe2:2023,PKim:2023,Huang:2024}. 
Given that the monolayer MoTe$_2$ sample in Ref.~\cite{xu_signatures_2025} exhibits a comparable mobility of $\mu \simeq 10^4$ cm$^2$/Vs, its disorder landscape is likely similar to other high-quality TMDs \cite{CDean_WSe2:2023,PKim:2023,Huang:2024}. Therefore, short-range disorder is expected to play a role in the emergence of the RIQAH2 state between $\nu = -2/3$ and $-3/5$, just as it does the analogous RIQH state in Ref.~\cite{li_observation_2010}.

Given the similarities between these experiments, we review the theoretical explanation of the RIQH offered in Refs.~\cite{Fogler:1996, Koulakov:1996}. They proposed a quantum Hall bubble phase that involves a mixture of a fraction $p_B$ of quasiparticles, say holes, on top of a filled electron LL (i.e. IQH liquid)~\cite{particle_hole}. These quasiparticles then form a bubble CDW or WC pinned by disorder to minimize their Coulomb energy \cite{WC}. 
Crucially, the proportion of the CDW/WC phase is determined by the \textit{magnetic filling fraction}, $\nu_{\mathrm{mag}} = 1-p_B$, rather than the filling fraction per lattice unit cell. See Fig.~\ref{fig:CDW}(a) for an illustration.

Similar physics may be at play in the RIQAH phase as well. This is substantiated by a recent density matrix renormalization group calculation which found a near-degeneracy in energies ($<1\%$) between FQAH, SC, and a RIQAH CDW phase in a realistic model of tMoTe$_2$~\cite{wang_chiral_2025}. This numerical calculation firmly places a CDW phase within the relevant phase competition \footnote{For a commensurate CDW, band folding may concentrate Berry curvature into a single subband, giving a topological CDW whose filling factor disperses with $B$ with integer slopes~\cite{xie_fractional_2021,Polshyn:2022}, which is different from the observed RIQAH.}. Motivated by the similarity of the RIQAH and RIQH effects, as well as this numerical work, we now generalize the physics of Refs.~\cite{Fogler:1996, Koulakov:1996} to include the presence of a background lattice. Suppose that we turn on a weak periodic potential, keeping the physics of the RIQH unaffected. 
We expect that this is a reasonable limit as long as the strength of the potential is less than the energy of the Wigner crystal. Since the potential strength is $\approx10$ meV in tMoTe$_2$ at $\theta\approx 3.9^{\circ}$~\cite{mao_transfer_2024, Duran:2024} while the energy of the Wigner crystal can be estimated in the jellium model to be on the order of tens of meV~\cite{WC_energy}, this is a reasonable approximation. Additionally, we note that in the Hofstadter model the small $\phi/\phi_0$ limit is dominated by LL physics and, as can be seen in Fig.~\ref{fig:streda_one}, the experiment is in this small flux limit. In the weak potential and small flux limit, to maintain a fixed magnetic filling $\nu_{\mathrm{mag}} = n h/eB$, the carrier density $n$ should change in proportion to the flux:
\begin{equation}\label{eq:nu_LL}
\nu(\phi) \equiv \frac{n(\phi)}{n_M} = \nu_{\mathrm{mag}}\frac{\phi}{\phi_0},
\end{equation}
where $\nu$ is the filling per unit cell, $n_M$ is the unit cell density, and $\phi=B/n_M$ is the flux per unit cell~\cite{LL_fan}. Thus a CDW occurring at $\nu_{\mathrm{mag}}=1-p_B$ will have this slope in a plot of $\nu$ versus $\phi$. Away from the weak periodic potential limit it may be more reasonable to expect any CDW to be formed due to the influence of the lattice potential, rather than interelectron repulsion. Thus the filling of the CDW may be fixed to the lattice filling rather than the magnetic filling factor, altering this slope. However, as mentioned, we expect the weak lattice limit to be reasonable for tMoTe$_2$ at $\theta\approx 3.9^{\circ}$.

To describe the RIQAH states seen in Ref.~\cite{xu_signatures_2025} it is necessary to make a few modifications to this approach. First one must account for the nonzero intercept $\nu(0)$ of Eq.~\eqref{eqn:Streda}. In a lattice system with an integer Chern band, the Berry curvature at $B = 0$ mimics a background magnetic field, which intercepts the conventional Landau fan at $1/C$ flux quantum. See Fig.~\ref{fig:CDW}(b) for an illustration. Second, one must account for the fact that the proximate IQAH state at $\nu=-1$ in Ref.~\cite{xu_signatures_2025} has a slope corresponding to a Chern number of $C=-1$. When this is done we find that a RIQAH state consisting of a fraction $p_B$ of quasiparticles forming a CDW/WC on top of the IQAH state at $\nu_{\mathrm{IQAH}}=-1$ will display the Streda formula
\begin{equation} \label{eqn:RIQAH_streda}
\nu(\phi) = (1-p_B)\left(C\frac{\phi}{\phi_0} + \nu_{\mathrm{IQAH}}\right).
\end{equation}
The above equation can be applied to the RIQAH state as a bubble or WC in a general Chern band with a Chern number $|C| \geq 1$. 
Thus this quantum Hall bubble or WC phase can manifest the fractional slope of the Streda formula. 

For the RIQAH phases of Ref.~\cite{xu_signatures_2025} we note that the proximate IQAH phase is located at $\nu_{\mathrm{IQAH}} = -1$ with $C=-1$. With these values Eq.~\eqref{eqn:RIQAH_streda} reveals that the slope and intercepts of any RIQAH phases should be equal. As reported by Ref.~\cite{xu_signatures_2025} this is satisfied by the RIQAH2 state.

We now comment on the effect of short-range disorder on the WC phase. Such disorder may lower the energy of the WC by slightly distorting the WC lattice position toward attractive impurities or away from repulsive impurities, which enhances the stability of WC + IQAH liquid mixture~\cite{joy_disorder-induced_2025,Xiang:2025,li_observation_2010}.
If this is the case for tMoTe$_2$ then we predict that: 1) making cleaner samples with fewer short-range impurities, 2) lowering the carrier density by lowering the twist angle, or 3) enhancing long-range disorder such as twist-angle disorder may all cause the RIQAH to disappear, just as the elimination of short-range disorder causes the disappearance of the RIQH state in the GaAs LLL~\cite{li_observation_2010}. 
Indeed, a similar RIQAH state was seen prior~\cite{Xiadong}.
We also note that if short-range disorder plays a role in the emergence of this RIQAH state it may complicate claims of anyon SC by introducing a few intermediate phase transitions that are not experimentally observed, as discussed in \cite{Shi:2025, Nosov:2025}. On the other hand, long-range disorder, such as twist-angle disorder and Coulomb disorder, also plays a role in localization and generating a finite plateau width \cite{Kazarinov:1982,Trugman:1983,Prange:1982,Joynt:1984, yi-thomas_integer_2025, Shi:2025, Nosov:2025}.

Having discussed the possible applicability of a LL bubble or WC picture to the situation where the ``Streda'' slope does not match the Hall conductivity as measured by transport, here we give some discussion of alternative explanations for different ``Streda'' slopes. We show that our proposed interpretation of short-range-disorder-enhanced Wigner crystal localization has an advantage over the alternative explanations.

First, we comment on the possibility of a stripe phase.
In a conventional stripe state, transport is metallic along the stripe direction (giving a $\rho_{xx}$ peak) but less conducting perpendicular to it (giving a $\rho_{xx}$ minimum). Tracking this $\rho_{xx}$ minimum versus $B$ and $n$ can therefore produce a slope resembling that of a ``Streda'' slope. 
However, this signal depends on the measurement axis being perpendicular to the stripe direction, which can be checked by rotating the current path and testing for resistivity anisotropy.
Moreover, the stripe phase is a metallic, compressible nematic with classical Hall conductivity $\sigma_{xy} = \nu e^2/h$ linear in the filling factor. If $\nu$ is noninteger, $\sigma_{xy}$ is not quantized, in sharp contrast with the integer-quantized Hall plateau observed in the RIQAH1 data.

There is a possible explanation relying on anyon dispersion in tMoTe$_2$. Without anyon dispersion, localization of fractional quasiparticles would yield a fractional rather than an integer Hall plateau. Reference~\cite{Nosov:2025} attributes the RIQAH to disorder lifting the degeneracy among multiple anyon-band minima, so that only one delocalized Landau-like state remains occupied. This explanation, however, crucially depends on anyon dispersion and therefore cannot account for the similar RIQH observed in GaAs, where no such dispersion exists. By contrast, our short-range-disorder-enhanced Wigner crystal localization scenario explains the RIQH in both tMoTe$_2$ and GaAs, offering a more universal and experimentally consistent framework.

Another option is a bubble phase with multiple fractional particles. A bubble phase with multiple fractional particles would correspond to a reentrant fractional quantum anomalous Hall (RFQAH) phase instead of the RIQAH phase. A similar reentrant fractional quantum Hall (RFQH) phase has been observed in high quality GaAs~\cite{huang2023reentrant,huang2024observation}. However, a RFQAH phase has not been observed experimentally, suggesting limitations of the bubble-phase interpretation and indicating that our short-range-disorder-enhanced Wigner crystal localization scenario provides a more consistent explanation.

Finally, we return to the mismatch of the Streda slope and intercept of the RIQAH1 state. It may be the case that the slope of the RIQAH1 state is misidentified due to its close proximity to the resistive peak displayed in Fig.~\ref{fig:streda_one}. Since the RIQAH phase is determined by finding the resistivity minima it can easily be overwhelmed when placed in close proximity to a resistive peak with a different slope. Indeed, we find the slope of the resistivity peak to correspond to a Chern number of $C=-0.49\pm 0.08$. This suggests the possibility that the RIQAH1 state is driven by the same physics as the RIQAH2 state, but its slope is misidentified due to its proximity to the dispersing peak in resistivity. We provide a line displaying what the slope of RIQAH1 would be if it was not masked by the resistive peak in Fig.~\ref{fig:streda_one}(c).

\section{VHS driven transition}
\label{sec:VHS}

What then is the origin of the resistivity peak that is found close by the RIQAH1 state? As discussed in the introduction and summarized in Fig.~\ref{fig:streda_one}, the peak appears to be related to a transition that changes the Hall conductivity, occurring around $\nu = -0.75$. The region of filling $\nu \in (-1, -0.75)$ with small Hall resistivity represents the largest difference between tMoTe$_2$ in Ref.~\cite{xu_signatures_2025} and GaAs LLL in  Ref.~\cite{li_observation_2010}. 
In Ref.~\cite{li_observation_2010} the IQH plateau is quite long, ranging all the way from $\nu_{\mathrm{mag}}=0.8$ to $\nu_{\mathrm{mag}}=1.4$, while in Ref.~\cite{xu_signatures_2025} the IQAH plateau in $\nu \in (-1.3,-1)$ is interrupted by this phase with small $\rho_{xy} \ll h/e^2$. This phase is consistent with observations in other tMoTe$_2$ samples with similar twist angles \cite{park_observation_2023,Xu:2023,Park:2025}.

The small anomalous Hall resistivity in the range $\nu \in (-1, -0.75)$ suggests that the intervening phase is partially valley-polarized. Such a valley-imbalanced state would be partially spin-polarized because of the locking of spin to valley in the TMD setting. Another possibility is that the state is IVC, analogous to similar phases in twisted WSe$_2$ \cite{fischer_theory_2024}. 

Whatever the nature of the phase, peaks in $\rho_{xx}$ and changes in $\rho_{xy}$ indicate that there are two transitions out of it, one near $\nu = -1$ and another near $\nu = -0.75$. We first focus on the transition near $\nu = -1$ before returning to our main interest at $\nu = -0.75$. Due to exchange interactions a spin-valley-polarized ferromagnetic phase is preferred near integer fillings.
Thus it is not surprising that the system tends to develop full spin-valley polarization around $\nu=-1$. On the other hand, away from integer fillings, metallic screening weakens exchange interactions, making the metallic phase with reduced polarization more favorable. As a theoretical reference, we therefore assume the phase in the range $\nu\in(-1,-0.75)$ to be a valley-unpolarized metal.

\begin{figure}
    \centering
    \includegraphics[width=\linewidth]{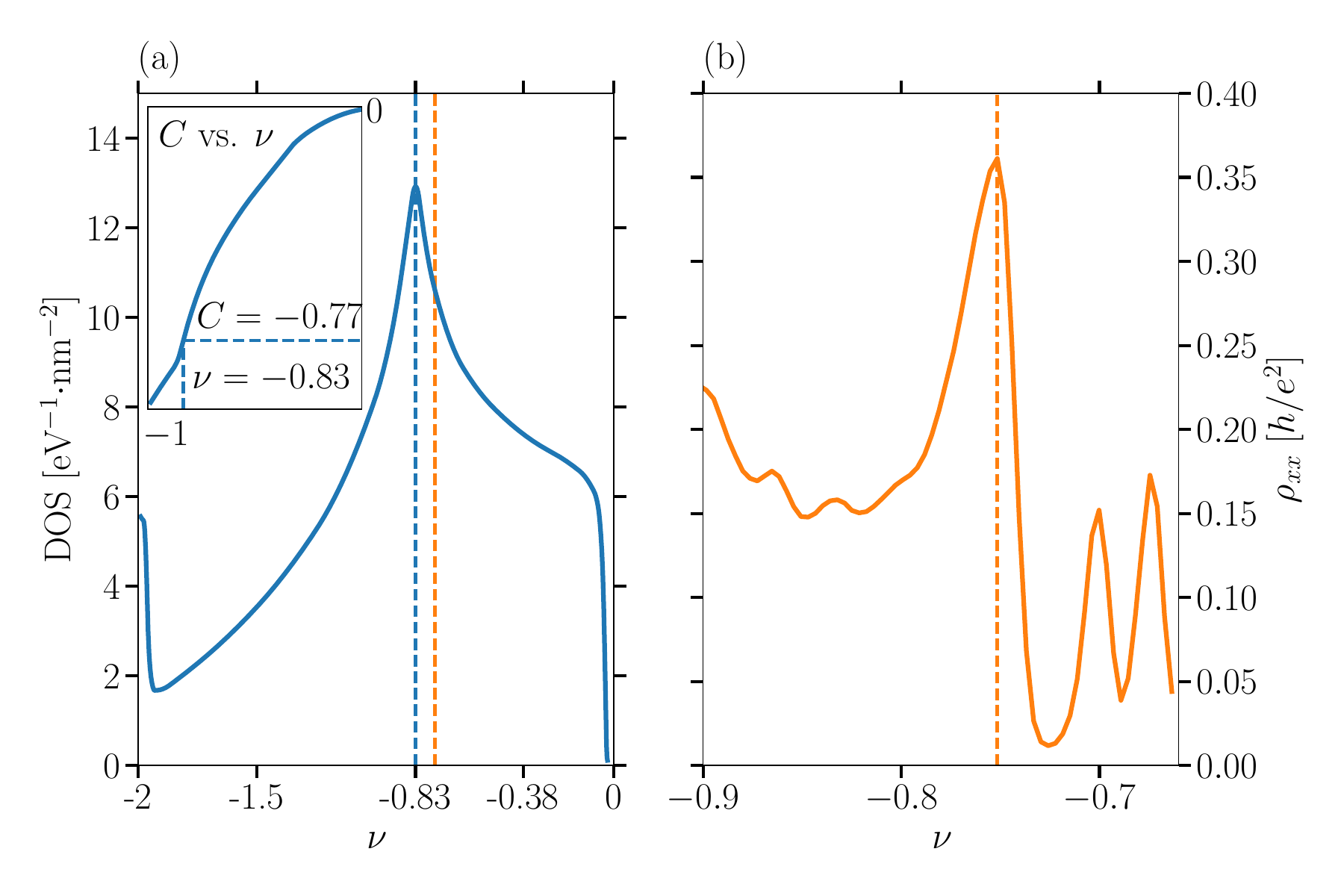}
    \caption{Single-particle density of states and experimental resistance as functions of doping, $\nu$. (a) The single particle density of states (DOS) for valley unpolarized tMoTe$_2$ at a twist angle of $3.8^{\circ}$. A DOS peak can be observed at the VHS at $\nu = -0.83$, indicated by the blue dotted line. Plotted in the inset is the single-particle Berry phase in units of $2\pi$ corresponding to a valley polarized Fermi surface. On the $x$ axis the filling runs from $-1$ to $0$ while on the $y$ axis $C$ runs from $-1$ to $0$. The value of $C$ is close to $-0.77$ near $\nu=-0.83$, indicated by the dotted lines. Also shown is the orange dotted line indicating the filling where the peak in resistivity appears, plotted more fully in (b). It can be seen to be close to the location of the VHS. (b) The experimental resistivity, $\rho_{xx}$, from Ref.~\cite{xu_signatures_2025} at $T=15$mK and $B=5$T is plotted in orange for $-0.9 <\nu <-0.66$. This is the same peak in resistivity highlighted in red markers in Fig.~\ref{fig:streda_one}(c). Note that the reason for choosing $B=5$T was to avoid the SC state that appears at $B=0$T.}
    \label{fig:DOS}
\end{figure}

We then calculate the single-particle density of states (DOS) using Eqs. (3) and (4) of Ref.~\cite{mao_transfer_2024} along with the continuum parameters there. We further make the assumption of a valley unpolarized metal. This is displayed in Fig.~\ref{fig:DOS}(a), where it can be seen that the DOS in the valley unpolarized metal exhibits a VHS at $\nu = -0.83$, potentially triggering a Stoner instability~\cite{Stoner:1938}. Figure~\ref{fig:DOS}(b) displays the experimentally observed resistivity from Ref.~\cite{xu_signatures_2025} at a high enough field to suppress the SC phase; the peak in the resistivity is indeed very close to the single-particle VHS as indicated by the dotted orange line on both subfigures. We note the VHS and the associated Fermi-surface topology change may also explain the experimentally observed sign reversal in $\rho_{xy}$~\cite{Kim:2016,Kokkinis:2022} across $\nu = -0.75$ at high magnetic fields~\cite{xu_signatures_2025}.

More precisely, the magnetic susceptibility takes the form $\chi \propto [1 - V(2k_F)\chi_0]^{-1}$, where $V(2k_F)=2\pi e^2/(\kappa 2k_F)$ is the Coulomb interaction, $\chi_0$ is the polarization function, and $k_F$ is the Fermi wave vector, both assuming two Fermi surfaces from two valleys~\cite{giuliani_quantum_2008,local_field}.
A magnetic instability occurs when $V(2k_F)\chi_0 > 1$, which is the Stoner criterion~\cite{Stoner:1938,giuliani_quantum_2008}.
Using $k_F=\sqrt{2\pi n} = \sqrt{2\pi \nu n_M}$, $\kappa = 5$ as the lattice dielectric constant for MoTe$_2$ encapsulated in hBN~\cite{Emanuel:2018,Laturia:2018}, and the moiré density $n_M=4.2\times 10^{12}$ cm$^{-2}$, we estimate $V(2k_F) = 1.8/\sqrt{\nu}\mathrm{\ eV \ nm}^{2}$.
The polarization function $\chi_0$ can be approximated by the single-particle DOS, which in two dimensions is $\mathrm{DOS} = m / \pi \hbar^2$ where we have taken into account the two fold degeneracy. 
Near the VHS, the DOS is enhanced to $\simeq 10\; \mathrm{eV}^{-1} \mathrm{nm}^{-2}$, see Fig.~\ref{fig:DOS}(a). This increases the effective mass to $m\simeq1.8 m_e$ (cf. the hole effective mass of monolayer MoTe$_2$ is $\simeq0.6m_e$~\cite{xu_signatures_2025}).
As a result, the dimensionless product near the VHS becomes $V\chi_0 \simeq 20$. 
The system is therefore highly susceptible to interaction-driven symmetry breaking, such as a first-order transition that spontaneously polarizes the valleys~\cite{Raines:2024a,Raines:2024b,Raines:2024c,Zhou_Xie_half_quarter_metal:2021,Zhou:2022,de_la_Barrera:2022,Arp:2024,Holleis:2025,Zhang:2023}.

Thus, the competition between kinetic and exchange energy near the VHS may drive a transition from the valley-imbalanced metal to a valley-polarized phase as the carrier density decreases, potentially explaining the resistive peak observed near this filling. There are at least two possible mechanisms that could drive this peak: First the peak could be due to a phase transition, as discussed earlier. If the order parameter couples to charged operators then near the critical point fluctuations of the order parameter will lead to stronger scattering and enhanced resistivity. Alternatively the peak could arise from the valley-polarized phase itself. This is because the valley-polarized state has weaker screening due to the missing degeneracy factor associated with valley degeneracy which suppresses electron polarizability. Suppressed screening in the valley-polarized state increases the effective impurity scattering strength, leading to an increased resistivity. This then leads to a peak in the resistivity, as has been discussed in the literature in the context of other 2D electron systems \cite{hossain_spontaneous_2021, ahn_valley_2022}. Such a resistivity peak is an indirect signature for a valley polarized metallic state.

Either explanation, however, leaves the question of why the resistive peak appears to disperse with magnetic field with a slope corresponding to a Chern number $C=-0.49 \pm 0.08$. This can be explained because it is a transition into a state with nonzero Chern number. In the quantum Hall context one example of this is a plateau transition between different IQH states. In the presence of a smooth, long-range disorder landscape, transport at the plateau transition can be understood as the percolation of domain boundaries separating regions with different Chern numbers (and thus different densities)~\cite{Wang:2014,Kazarinov:1982,Trugman:1983,Prange:1982,Joynt:1984}. Thus the transition can be expected to disperse with the magnetic field. Indeed, when the Berry phase (in units of $2\pi$) on the valley-polarized side at $\nu = -0.83$ is computed, it is found to be $-0.77$, see the inset of Fig.~\ref{fig:DOS}(a). So it is perhaps not surprising that the transition appears to disperse at a slope close to this value.

\section{Discussion}
\label{sec:disc}

In this paper we provide a physical scenario for the fascinating experiment described in Ref.~\cite{xu_signatures_2025}. In particular we focused on the appearance of the RIQAH states whose Hall conductivity extracted via the Streda formula does not agree with the Hall conductivity as measured by transport. We showed that one possible reason this may happen is if the system has a localized compressible state analogous to the classic LL bubble or WC picture. If this is the case, the RIQAH2 state between $\nu = -2/3$ and $-3/5$ can be understood as a WC built from localized charge-1 quasiparticles counted relative to the $\nu = -1$ IQAH state. The adjacent $\rho_{xx}$ peaks then mark the boundaries where the system switches from charge-1 localization to localization of the fractional quasiparticles that underpin the neighboring $\nu = -2/3$ and $-3/5$ plateaus. This is in contrast with the scenario proposed in Ref.~\cite{Nosov:2025} where the RIQAH state arises from these fractional quasiparticles.

We then discussed the similarities between the RIQAH states and the RIQH states in the continuum quantum Hall setting in GaAs LLL~\cite{li_observation_2010}.  In particular, we highlighted the likely importance of short-range disorder for both states and noted that the RIQH state disappears when short-range disorder is reduced. To experimentally test the electron solid scenario, thermodynamic compressibility and noise in nonlinear transport can be probed, as demonstrated in Refs.~\cite{Young:2023,zeng_thermodynamic_2023,Bennaceur:2018}. While the bubble or WC picture is able to straightforwardly explain the RIQAH2 state seen in Ref.~\cite{xu_signatures_2025}, it cannot immediately explain the RIQAH1 state. However, the resistivity peak adjacent to this RIQAH1 state may disguise its intrinsic slope.

We thus focused on this resistivity peak around $\nu = -0.75$ and noted that it appears to coincide with a transition from nearly quantized $\rho_{xy} = h/e^2$ to a small $\rho_{xy} \ll h/e^2$, and the latter is a valley-imbalanced metal. We took the simplifying assumption that the valley imbalanced phase is valley unpolarized and computed the single-particle density of states. We found a VHS that occurred at $\nu=-0.83$, near $\nu = -0.75$, which may drive the transition via mechanisms such as the Stoner instability. The appearance of a VHS that drives this transition could also explain the change of sign of $\rho_{xy}$ across $\nu = -0.75$ at large magnetic fields. We note that such a transition from valley imbalanced to valley fully polarized phase may be a first-order Stoner transition~\cite{Raines:2024a,Raines:2024b,Raines:2024c,Zhou_Xie_half_quarter_metal:2021,Zhou:2022,de_la_Barrera:2022,Arp:2024,Holleis:2025,Zhang:2023}. A signature of this abrupt change in the size of the Fermi surface would be a discontinuity in quantum oscillations. Although not accessible with current devices, future cleaner generations of tMoTe$_2$ may be able to probe this possibility.

We mention another possibility is an insulator-metal-insulator transition (driven by disorder-induced localization transition) where a fully valley-polarized anomalous Hall metal~\cite{Crepel:2023} appears in between QAH insulators. In this scenario, the bulk remains a compressible conducting liquid due to screened disorder in the high filling range $\nu \in (-1, -0.75)$. However, localization is expected to be stronger near the single-valley band edge at $\nu = -1$, which is inconsistent with the observed metallic behavior there--making this interpretation less compelling.

We comment on the recently observed extended quantum anomalous Hall (EQAH) effect in rhombohedral-graphene moir\'e systems~\cite{lu2024extendedquantumanomaloushall}. Extended Fig.~9 of Ref.~\cite{lu2024extendedquantumanomaloushall} shows a RIQAH plateau between fillings $2/3$ and $3/5$ at low temperature, closely resembling the tMoTe$_2$ data in Ref.\cite{xu_signatures_2025}. This parallel phenomenology across two very different platforms may imply a common underlying mechanism. Moreover, at sufficiently low temperature and bias the EQAH state competes with nearby FQAH states and extends over a broader range of moir\'e fillings. A natural interpretation is that a WC/CDW is the lowest-energy state at ultralow temperatures --- even at fractional fillings, so observing FQAH requires melting the WC by gently raising the temperature while remaining below the FQAH melting point. Similar FQH emerges from melting of WC has been observed in GaAs~\cite{Pan:2002}.

As noted in the introduction, the SC observed near $\nu = -0.75$ in Ref.~\cite{xu_signatures_2025} may be linked to a nearby VHS-driven transition. This may suggest a SC emerging out of a valley imbalanced normal state via more conventional pairing mechanisms, e.g., BCS-like pairings \cite{chou_acoustic-phonon-mediated_2021,zhu_superconductivity_2025}, Kohn-Luttinger \cite{ghazaryan_unconventional_2021, jimeno-pozo_superconductivity_2023, guerci_topological_2024, qin_kohn-luttinger_2024}, and antiferromagnetic spin fluctuations \cite{fischer_theory_2024}.
This is an interesting direction for future work.
Though we do not investigate it further here we offer a comment on one alternative explanation of this SC as an anyon SC \cite{shi_doping_2024, Shi:2025, Nosov:2025, pichler_microscopic_2025}. As discussed earlier in Sec.~\ref{sec:mismatch} short-range disorder appears to be the dominant source of scattering in high-quality flux-grown TMDs, which likely stabilizes the RIQAH state.
However, for the anyon SC short-range disorder introduces intermediate transitions not seen experimentally \cite{Shi:2025, Nosov:2025}. We suggest probing the same region of phase space in samples with controlled increases in short-range disorder: if the SC-FQAH boundary becomes more robust, it would support a more conventional pairing mechanism.

Finally, we comment on an alternative explanation of the SC in Ref.~\cite{xu_signatures_2025} as a chiral topological superconductor which can host Majorana edge states \cite{xu_chiral_2025}. The same SC has been proposed in Refs.~\cite{geier_chiral_2024, chou_intravalley_2025, wang_chiral_2024, yang_topological_2024, parra-martinez_band_2025, christos_finite-momentum_2025,Dong:2025,Gil:2025,Jahin:2025,Qin:2024,Yoon:2025} for the SC1 in the electron-doped rhombohedral multilayer graphene (RMG)~\cite{Han:2025}. However, under a perpendicular magnetic field, the phase boundary between RMG SC1 and the quarter metal phase remains unchanged or even broadens, whereas the SC phase in tMoTe$_2$ contracts with an increasing magnetic field, suggesting a distinct phenomenology between the two SCs. We note that the out-of-plane magnetic field response for SC in tMoTe$_2$ is qualitatively similar to SC2 of RMG, which has been predicted to be topologically trivial due to annular Fermi surfaces \cite{chou_intravalley_2025,geier_chiral_2024,yang_topological_2024}. The nature of the SC observed in Ref.~\cite{xu_signatures_2025} is very much an open question, and one that we hope will be clarified by future experiments. 

\acknowledgements 
The authors acknowledge helpful conversations with Jiaqi Cai, Long Ju, Pavel Nosov, Jay D. Sau, T. Senthil, Qianhui Shi, Zhengyan Darius Shi, Boris Shklovskii, Xiaodong Xu, and Michael Zudov.
Y.H., S.M., and J.Z. thank Tingxin Li for helpful discussions and for sharing data.
The authors were supported by the Laboratory for Physical Sciences.

\appendix

\section{Line fitting to resistivity minima/maxima}
\label{app:line_fitting}

In this appendix we provide a brief description of how the lines were fit to the resistivity minima and maxima in Fig.~\ref{fig:streda_one}.

\begin{figure}
    \centering
    \includegraphics[width=\columnwidth]{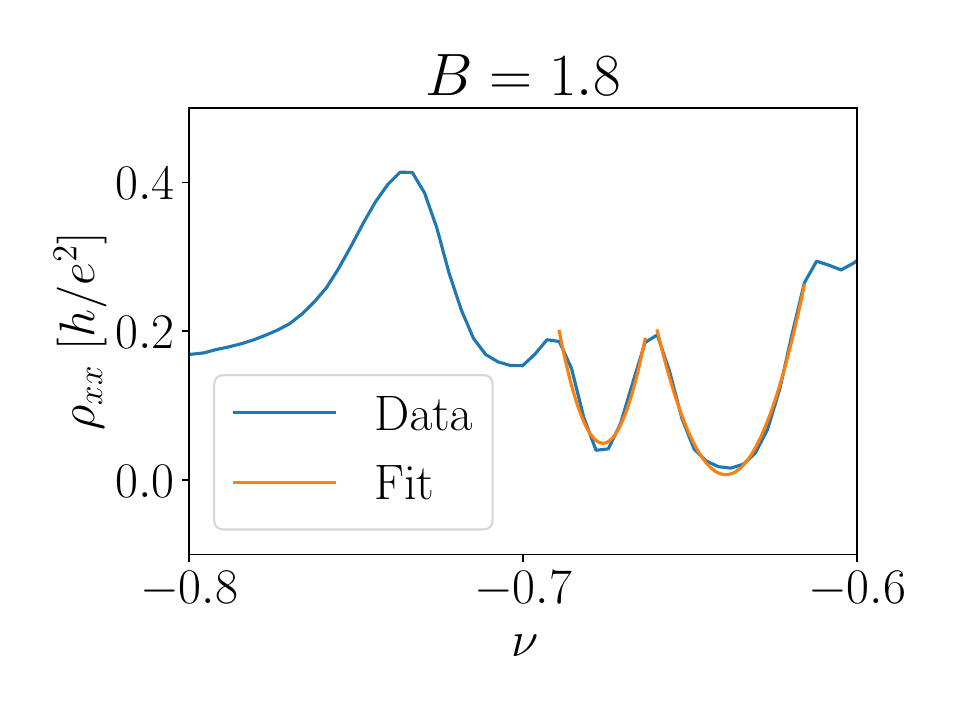}
    \caption{A plot of the fitting process used to avoid noise in the resistivity minima and maxima. Once candidate minima and maxima were identified parabolas were fit to smooth out any noise near the minima. The plot displays fits of parabolas to two identified minima at $B=1.8$. The leftmost minima corresponds to the FQAH state at $\nu = -2/3$ while the rightmost corresponds to the RIQAH2 state.}
    \label{fig:parabola_fits}
\end{figure}

First the functions $\verb|LassoSelector|$ and $\verb|Path|$ from $\verb|matplotlib|$ were used to manually select the regions of the data to consider. This was done to avoid the SC dome displayed in Fig.~\ref{fig:streda_one}. Then the $\verb|argrelextrema|$ function in $\verb|scipy|$ was used to detect local minima and maxima of $\rho_{xx}$ versus $\nu$ for each value of $B$. These were temporarily stored as candidate minima and maxima, provided they fell in the selected region. To eliminate noise in the data a parabola in $\rho_{xx}$ versus $\nu$ was then fit seeded from each candidate value. The fitting was done by the $\verb|curve_fit|$ function from $\verb|scipy|$. The parameters of the parabola and the number of points in the fit were chosen dynamically to minimize the reduced $\chi^2$ of the fit. The location of the minimum and maximum of the best-fit parabola was accepted as the final minimum and maximum value. Figure~\ref{fig:parabola_fits} displays a plot of $\rho_{xx}$ versus $\nu$ at $B=1.8$ as an example to illustrate this procedure.

Once the minima and maxima were finalized they were grouped into lines in $B$ versus $\nu$. The function $\verb|HoughLinesPointSet|$ from $\verb|OpenCV|$ was used to detect candidate lines through the data, where we took $|B|$ so data collapsed onto lines with the same slope. The distances of each point to each of these candidate lines were calculated and stored as a 2D array. Finally the feature detection function $\verb|HDBSCAN|$ from $\verb|sklearn|$ was used to separate the point groupings into their respective lines. Only points with a probability of $10\%$ or greater of belonging to the same line were accepted, otherwise they were discarded.

Once the minima and maxima were separated into their respective lines $\verb|curve_fit|$ from $\verb|scipy|$ was used to fit a line to each candidate. Following Eq.~\ref{eqn:RIQAH_streda} with $C=-1, \nu_{\mathrm{IQAH}}=-1$ the slope and intercept of the fitting lines through $\rho_{xx}$ minima were taken to be equal to one another. Since there is no theoretical reason for the slope and intercept of the lines through the maxima to agree, the slope and intercept of the fitting lines through the maxima were allowed to disagree. Note that because there were more data points for $B>0$ the best-fit lines are naturally weighted more toward $B>0$.

After performing this fit the data points, with their error, were plotted on top of the resistivity data. On top of these data points were then plotted the best-fit lines.

\bibliographystyle{apsrev4-2_custom}
\bibliography{tMoTe2.bib}

\begin{thebibliography}{115}
\bibitem [{mas()}]{mask}%
  \BibitemOpen
  \href@noop {} {}\bibinfo {note} {The $\rho_{xx}$-minimum intercept at $B=0$ might also be obscured by a neighboring $\rho_{xx}$ peak. Then a theoretical prediction for the ``true'' slope would need details of that peak’s nature and further measurements.}\BibitemShut {Stop}%
\bibitem [{par()}]{particle_hole}%
  \BibitemOpen
  \href@noop {} {}\bibinfo {note} {The quasiparticles could be electrons or holes for a LL, due to particle hole symmetry. The particle hole symmetry is broken by the finite kinetic energy for a Chern band in a moiré system.}\BibitemShut {Stop}%
\bibitem [{WC()}]{WC}%
  \BibitemOpen
  \href@noop {} {}\bibinfo {note} {Experimental support for this view is provied by the fact that when the IQH liquid background is removed from the RIQH state at $\nu_{\mathrm{mag}} \in (3/5, 2/3)$ in the GaAs LLL, the particle-hole counterpart of the state at $\nu_{\mathrm{mag}} \in (1/3, 2/5)$ exhibits insulating behavior, with $\rho_{xx}$ increasing as temperature decreases~\cite{li_observation_2010}, consistent with a pinned WC scenario.}\BibitemShut {Stop}%
\end{thebibliography}

\end{document}